\documentclass[12pt]{article}
\usepackage{amsmath,amssymb,amsfonts,epsf}
\usepackage[nosort]{cite}
\usepackage[margin=1in]{geometry}

\usepackage{fleqn}

\begin{document}

\begin{center}\ \\ \vspace{60pt}
{\Large {\bf Baryons in the Field Correlator Method}}\\ 
\vspace{30pt}
Roman Ya. Kezerashvili$^{1,2}$, I. M. Narodetskii$^{1,3}$ and A. I. Veselov$^{3}$
\vspace{20pt}

{\it $^1$Physics Department, New York City College of Technology,\\ The City University of New York, Brooklyn, NY 11201, USA.}

\vspace{10pt}
{\it $^2$The Graduate School and University Center,\\ The City University of New York, New York, NY 10016, USA}\\

\vspace{10pt}
{\it $^3$Institute of Theoretical and Experimental Physics, Moscow 117218, RF}\\

\end{center}

\vspace{30pt}

\centerline{\bf Abstract}

\noindent The ground and $P$-wave excited states of $nnn$, $nns$ and $ssn$
baryons are studied in the framework of the field correlator
method using the  running strong coupling constant in the
Coulomb--like part of the three--quark potential.
The string correction
for the confinement potential of the orbitally excited baryons,
which is the leading contribution of the proper inertia of the
rotating strings, is estimated.

\vspace{10pt}

\noindent {\bf keywords:}\ \ Baryons spectroscopy, S- and P-wave baryons, strong coupling constant

\thispagestyle{empty}
\vspace{10pt}

\section{INTRODUCTION}

Quantum chromodynamics (QCD) has been established as the theory describing
the strong interaction but its application to low--energy hadron
phenomenology is still far from a routine deduction. Various approximations,
whose connection to the underlying theory remains sometimes obscure, are
presently used to describe baryon spectroscopy. 
It has become an attractive program to develop model independent methods
which are firmly based in fundamental theory. 

The field correlator method (FCM) in QCD \cite{DS} provides promising
perspective. The FCM is a formulation of the nonperturbative QCD that gives
additional support to the quark model assumptions. Progress was recently
made \cite{NSV,AF} towards placing the computation of baryon masses
within the FCM on the same level as that of mesons. Nevertheless, this work
can be refined. In Refs. \cite{NSV} a freezing value of the strong coupling
constant in the perturbative Coulomb--like potential has been employed. This
choice appears to be a reasonable approximation and gives rise to a good
description of heavy quarkonia and heavy--light mesons. Note that 
for light baryons the Coulomb--like force does not play a crucial role and
produces only a marginal ($\sim 10\%$) correction. Nevertheless, it is
important to include into the FCM approach the modern knowledge about the
one--gluon exchange forces that still represent a fundamental concept which
might give us a deeper understanding of baryon spectroscopy.

In this talk, we explore the FCM for baryons in a more regular way by
considering the effects of the running strong coupling constant in the
Coulomb--like part of the three--quark potential. We use the background
perturbation theory (BPTh) \cite{Simonov1995} for the coupling constant to
avoid the infrared singularities. Below are considered the baryons composed
of valence light quarks, the \textit{up, down,} and \textit{strange}
flavors. We present the new results for the masses of the ground states and $%
P$--wave excited states of $nnn$, $nns$ and $ssn$ baryons. We also estimate
the string correction to the baryonic orbital excitations.

\section{EFFECTIVE HAMILTONIAN in FCM}

In the FCM three--quark dynamics in a baryon is encoded in gluonic field
correlators which are responsible for quark confinement. Starting from the
Feynman--Schwinger representation for the quark and gluon propagators in the
external field, one can extract hadronic Green's functions and calculate the
baryon spectra. 
The key ingredient of the FCM is the use of the auxiliary fields (AF)
initially introduced in order to get rid of the square roots appearing in
the relativistic Hamiltonian \cite{AF}. 
Using the AF formalism allows one to derive a simple local form of the
effective Hamiltonian (EH) for the three-quark system, 
which comprises both confinement and relativistic effects, and contains only
universal parameters: the string tension $\sigma$, the strong coupling
constant $\alpha_s$, and the bare (current) quark masses $m_i$.

The baryon mass is given by 
\begin{equation}  \label{M_B}
M_B\,=\,M_0\,+\,\Delta M_{\mathrm{string}}\,+\,C,\,\,\,\,\,\,
M_0\,=\,\sum\limits_{i=1}^3\left(\frac {m_{i}^2}{2\mu_i\,}+ \,\frac{\mu_i}{2}%
\right)\,+\,E_0(\mu_i)
\end{equation}
where $E_0(\mu_i)$ is an eigenvalue of the Schr\"{o}dinger operator $H_0 +V$%
, $V$ is the sum of the string potential $V_Y(\mathbf{r}_1,\,\mathbf{r}_2,\,%
\mathbf{r}_3)\,=\,\sigma\,r_{min}$, $r_{min}$ being the minimal string
length corresponding to the Y--shaped configuration,\footnote{%
This potential naturally arises from the consideration of the Wilson loop
depicted in Fig. 1a.} and a Coulomb interaction term $V_{\mathrm{C}}(\mathbf{r}%
_1,\,\mathbf{r}_2,\,\mathbf{r}_3)\,=\,\sum_{i<j}\,V_C(r_{ij})$, arising from
the one-gluon exchange. The constant auxiliary fields $\mu_i$ finally
acquire the meaning of the dynamical quark masses. They are defined by
minimization condition. 
The quark self-energy correction $C$ that is created by the color magnetic
moment of a quark propagating through the vacuum background field. This
correction, which can be calculated perturbatively, adds an overall negative
constant to the hadron masses \cite{S2001}.

Note that the string potential $V_Y(\mathbf{r}_1,\,\mathbf{r}_2,\,\mathbf{r}%
_3)$ represents only the leading term in the expansion of the QCD string
Hamiltonian in terms of angular velocities \cite{DKS}. The leading
correction in this expansion is known as a string correction. This is the
correction totally missing in relativistic equations with local potentials.
Its sign is negative, so the contribution of the string correction lowers
the energy of the system, thus giving a negative contribution to the masses
of orbitally excited states, leaving the S--wave states intact.

The calculation of the string correction is greatly simplified, if the
string junction point is chosen to coincide with the center--of--mass
coordinate ${\boldsymbol{R}}_{\text cm}$. In this case, the complicated
string junction potential is approximated by a sum of the one--body
confining potentials. The accuracy of this approximation for the $P$--wave
baryon states is better than 1$\%$ \cite{AF}. As the result one obtains 
\begin{equation}  \label{eq:Delta_M}
\Delta M_{\mathrm{string}}\,=\,-\,\frac{\sigma}{6}\,<\Psi|\sum_i\,\frac{({%
\boldsymbol{r}}_i\times{\boldsymbol{p}}_i)^2}{\mu_i^2\,r_i}\,|\Psi>,
\end{equation}
where $\Psi$ is the eigenfunction of the unperturbed EH. 
%
%

\section{COULOMB-LIKE INTERACTION in the BPTh}

Details of our treatment of the string junction confinement potential can be
found in Ref. \cite{NSV}. We now concentrate on the Coulomb--like part of
interaction, $V_C(r)$. It is convenient to write the Coulomb--like potential
in QCD in momentum space as 
\begin{equation}  \label{eq:MS}
V_C({\mathbf{q}}^2)\,=\,-\,C_F\,\frac{\alpha_V({\mathbf{q}}^2)}{{\mathbf{q}}%
^2},
\end{equation}
where $C_F$ is the color factor. 
A running constant $\alpha_V({\mathbf{q}}^2)$ controls the behavior of
standard perturbation theory (SPTh) at low momentum scales. The formal
expression for $V_C(r)$ in position space is written as 
\begin{equation}  \label{eq:position space}
V_C(r)\,=\,-\,C_F\,\frac{\alpha_s(r)}{r}\,,\,\,\,\,\,\, \alpha_s(r)\,=\,%
\frac{2}{\pi}\,\int\limits_0^{\infty}dq\,\,\frac{\sin\,qr}{q}\,\,\alpha_V({%
\mathbf{q}}^2).
\end{equation}
Up to two loops 
\begin{equation}  \label{eq:alpha_V}
\alpha_V({\mathbf{q}}^2)\,=\,\frac{4\pi}{\beta_0\,t}\left(1\,-\,\frac{\beta_1%
}{\beta_0\,^2}\,\frac{\ln t}{t}\right),\,\,\,\,\,\,\,\,\,\,\,t\,=\,\frac{%
\boldsymbol{q}^2}{\Lambda_V^2}\,,
\end{equation}
where $\beta_i$ are the coefficients of the QCD $\beta$-function. 

The conventional coupling (\ref{eq:alpha_V}) is analytically singular at a
scale ${\mathbf{q}}^2\,=\,\Lambda_{V}^2$, so the SPTh itself is not well
defined in the infrared domain where the coupling become large. This problem
can be traced back to the fact that the integral over the running coupling
which appears in Eq.(\ref{eq:position space}) is ill defined. 
As a result, $\alpha_s(r)$ is known only in the perturbative region, $%
r\,\lesssim\, 0.1$ fm. Estimates of the average interquark distances in
light baryons are in the vicinity of 0.7 fm which is certainly outside of
the perturbative region.

There also exist the possibilities of defining a running coupling which
stays finite in the infrared. For our purposes, we find it convenient, as a
useful approximation, to define the strong coupling constant $\alpha_B(r)$
in the BPTh \cite{Simonov1995}. The logic behind this approach is that the
perturbative gluon propagator is modified strongly at $q\,\lesssim m_B$ by
the physics of large distances. In momentum space, $\alpha_B({\mathbf{q}}%
^2)\,=\,\alpha_s({\mathbf{q}}^2\,+\,m_B^2)$, where $m_B\,\sim\,1$ Gev has
the meaning of the lowest hybrid excitation; from comparison with the
lattice static potential $m_B\,\sim\,1\,$ GeV. The result can be
conventionally viewed as arising from the interaction of a gluon with
background vacuum fields.

We define $\alpha_B(r)$, as well as the Coulomb--like potential in the
configuration space via expressions similar to (\ref{eq:position space})
with the substitution $\alpha_s(\boldsymbol{q}^2)\rightarrow\alpha_B(%
\boldsymbol{q}^2)$ and $\alpha_s(r)\rightarrow\alpha_B(r)$. 
For $\alpha_B({\mathbf{q}}^2)$, we use the standard two--loop result with
the substitution $t\,\rightarrow\,t_B\,=\,\ln\,\frac{{\mathbf{q}}^2\,+\,m_B^2%
}{\Lambda_V^2}$. The resulting coupling $\alpha_B({\mathbf{q}}^2)$ is finite
in the infrared, ${\mathbf{q}}^2\to 0$. In the ultraviolet region, ${\mathbf{%
q}}^2\gg m_B^2$ one recovers the SPTh result. In configuration space the
background coupling constant $\alpha_B(r)$ exists for all distances and
saturates at some critical, or freezing, value for $r\,\gg\, 1/m_B$. 

\section{RESULTS and DISCUSSION}

\label{sect:results} 

\begin{figure}[ht]
   \epsfxsize=6.0in \centerline{\epsffile{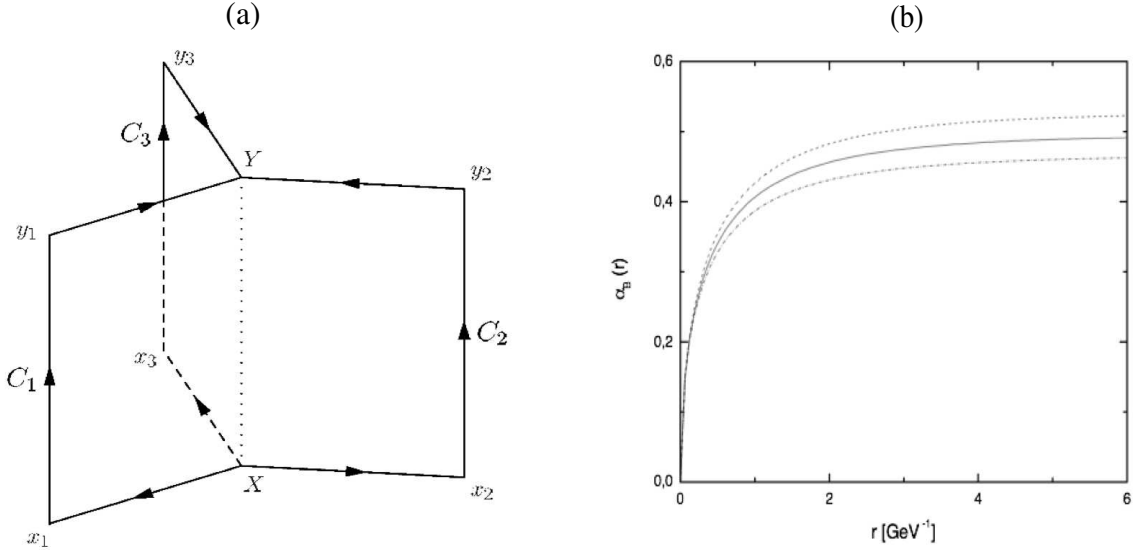}}
   \caption[FIG. \arabic{figure}.]{\footnotesize{ (a) Three--lobes Wilson loop for a baryon. 
(b)  The dependence of the running coupling constant $\protect\alpha%
_B(r) $ on the distance for $m_B\,=\,$1 GeV and $\Lambda_V\,=\,$ 0.36 (solid
line), 0.38 (dashed line) and 0.34 GeV (dotted--dashed line)}}
\label{fig:freezing}
\end{figure}

The EH contains five parameters: the current quark
masses $m_{n}$ and $m_{s}$, the string tension $\sigma $, and two parameters 
$\Lambda _{V}$ and $m_{B}$ defining $\alpha _{B}(r)$. Let us underline that
they are not the fitting parameters. In our calculations we use $\sigma
\,=\, $0.15 GeV$^{2}$ found in the SU(3) QCD lattice simulations. We employ
the current light quark masses ${m}_{u}\,=\,{\ m}_{d}\,=\,7\,$ MeV and the
bare strange quark mass $m_{s}\,=\,$ 175 MeV found previously from the fit
to $D_{s}$ spectra. This value of the strange quark mass is consistent with
the QCD sum rules estimation $m_{s}\,$(2~GeV)$\,=\,(125\pm 40)$ MeV. 
much lower scale. For the remaining two parameters $\Lambda _{V}$ and $m_{B}$
we employ the values $\Lambda _{V}\,=\,(0.36\pm 0.02)\,\mathrm{GeV}%
,\,m_{B}\,=\,(1\pm 0.05)\,\mathrm{GeV},$ determined previously in Ref. \cite%
{BK2001}. The result is consistent with the freezing of $\alpha _{B}(r)$ at $%
r\gg m_{B}^{-1}$ with a magnitude $\sim \,0.5-0.6$. The behavior of $\alpha
_{B}(r)$ for $\Lambda _{V}\,=\,0.36$ GeV is shown in Fig. \ref{fig:freezing}b
for three different values of $m_{B}$. 
Note that $\alpha _{B}(r)$ increases with $\Lambda _{V}$ and, for fixed $%
\Lambda _{V}$, decreases with $m_{B}$.

We begin the discussion of our results\footnote{%
For more detailed presentation of the results see Ref. \cite{KNV}} by examining our
predictions for the ground states of the $nnn$, $nns$ and $snn$ baryon with $%
L\,=\,0$. 
First, we mention that the baryon masses decrease when $\Lambda_V$ increases
and, for fixed $\Lambda_V$, the baryon masses decrease when $m_B$ increases.
This comes as no surprise: the effect can be easily read off from 
Fig. \ref{fig:freezing}b. Increasing $\Lambda_V$ for fixed $m_B$ and
decreasing $m_B$ for fixed $\Lambda_V$ leads to increased running constant $%
\alpha_B(r)$ and respectively, to decreased baryon masses. Upon varying the
parameters $\Lambda_V$ and $m_B$ within the error bars indicated above, we
obtain the baryon masses in the interval 1161\,--\,1209 MeV ($nnn$),
1246\,--\,1297 MeV ($nns$), and 1330\,--\,1383 MeV ($ssn$). The difference
in the mass values is mostly due to the difference in the running of the
strong coupling in the midmomentum regime.

The baryon energies agree reasonably well with the Particle Data Group (PDG)
listing particularly if we take into consideration that spin interactions
are neglected. For instance, for $L\,=\,0$ we get $\frac{1}{4}%
(\Lambda\,+\,\Sigma\,+\,2\,\Sigma^*)_{\mathrm{theory}}\,=\,$ 1276$%
^{+23}_{-30}$ MeV, where the error bars correspond to the variation of the
hyperon masses within the chosen range of $\Lambda_V$ and $m_B$ versus $%
\frac{1}{4}(\Lambda\,+\,\Sigma\,+\,2\,\Sigma^*)_{\mathrm{exp}}\,=\,$ 1267
MeV. For the $\Xi$, we have $\Xi_{\mathrm{theory}}\,=\,$1360$^{+23}_{-30}$
MeV, whereas $\Xi_{\mathrm{exp}}\,=\,$1315 MeV. However, for the nucleon we
get $\frac{1}{2}\,(N\,+\,\Delta)_{\mathrm{theory}}\,=\,$ 1187$^{+22}_{-21}$
MeV, which is 
about 100 MeV heavier than $\frac{1}{2}\,(N\,+\,\Delta)_{\mathrm{exp}}\,=\,$
1085 MeV. The difference can be ascribed to the effects of spin--dependent
quark--quark interactions modeled after the effect of gluon exchange in QCD
or arising from one--boson exchange that are completely omitted in the
present approach. Another source of the discrepancy is the systematic error
associated with the use of the AF formalism, which is maximal for the $S$%
--wave $nnn$ states \cite{AF}.

Analysis of the P-wave excitations shows that string correction does not
essentially depend on the baryon flavor nor on the type of excitation. As a
result of the weak dependence of the string correction on a baryon flavor
and the type of excitation the masses for all considered baryons become
smaller by about the same value $\sim\,50-60$ MeV.

The physical $P$-wave states are not pure $\rho$ or $\lambda$ excitations
but linear combinations of all states with a given total momentum $J$. Most
physical states are, however, close to pure $\rho$ or $\lambda$ states. 
 For example, the masses of N(1535) and N(1520) resonances with $J^P\,=\,%
\frac{1}{2}^-$ and $J^P\,=\,\frac{3}{2}^-$, respectively, both match with
the mass of the $\lambda$--excitation for the $nnn$ baryon: $%
M_{\lambda}(nnn)\,=\,1567^{+13}_{-14}$ MeV. The masses of $\Sigma(1620)$ and 
$\Sigma(1670)$ states with $J^P\,=\,\frac{1}{2}^-$ and $J^P\,=\,\frac{3}{2}%
^- $, respectively, match very closely with the mass of the $\lambda$%
--excitation for the $nns$ baryon: $M_{\lambda}(nns)\,=\,1636^{+13}_{-15}$
MeV. The masses of the $\rho$--excitations which correspond to $P$--states
of the light diquarks are typically 60 - 80 MeV higher.

\section{CONCLUSIONS}

In this paper the study of the ground and excited states of the $qqq$, $qqs$
and $ssq$ baryons by including into the EH derived using the FCM the effects
of the running strong coupling constant $\alpha_B(r_{ij})$ in the
perturbative Coulomb--like part of the three--quark potential is presented.
The results have refined our previous studies for the ground and excited
states of the $qqq$, $qqs$ and $ssq$ baryons obtained for the freezing
coupling constant. 
Our study shows that a fairly good description of the $S$ and $P$--wave
baryons can be obtained with spin independent energy eigenvalues
corresponding to the confining along with Coulomb potentials. We emphasize
that no fitting parameter were used in our calculations. This comparative
study provides deeper insight into the quark model results for which the
constituent masses encode the QCD dynamics.

\section*{Acknowledgements} This work was supported in part by RFBR Grants 08-02-00657,
08-02-00677, 09-02-00629 and by the grant for scientific schools
NSh.4961.2008.2.







\begin{thebibliography}{9}

\bibitem{DS} For a review see: A.~Di Giacomo, H.~G.~Dosch, V.~I.~Shevchenko,
Yu.~A.~Simonov, \emph{Phys. Rep.} \textbf{372}, 319 (2002).

\bibitem{NSV} I. M.~Narodetskii and M. A.~Trusov, \emph{Phys.~Atom.~Nucl.} \textbf{67}, 762 (2004); O.~N.~Driga, I.~M.~Narodetskii, A.~I.~Veselov, \emph{Phys. Atom. Nucl.} \textbf{71}, 335 (2008)

\bibitem{AF} I.~M.~Narodetskii, C.~Semay, A.~I.~Veselov, \emph{Eur. Phys. J. C} 
\textbf{55}, 403 (2008).

\bibitem{Simonov1995} Yu.~A.~Simonov, \emph{Phys. Atom. Nucl.} \textbf{58}, 107
(1995) [\emph{Yad. Fiz.} \textbf{58}, 113 (1995)].

\bibitem{DKS} A. Yu. Dubin, A. B. Kaidalov and Yu. A. Simonov, \emph{Phys. Lett. B} 
\textbf{323}, 41 (1994); A. Yu. Dubin, A. B. Kaidalov and Yu.A. Simonov,
\emph{Phys. Atom. Nucl.} \textbf{56}, 1745 (1993).

%
%

\bibitem{S2001} Yu.~A.~Simonov, \emph{Phys. Lett. B} \textbf{515}, 137 (2001). 

\bibitem{BK2001} A.~M.~Badalian, \emph{Phys. At. Nuclei} \textbf{63}, 2173 (2001),
A.~M.~Badalian and D.~S.~Kuzmenko, \emph{Phys. Rev. D} \textbf{65}, 016004 (2002).

\bibitem{KNV} R.~Ya.~Kezerashvili, I.~M.~Narodetskiy, A.~I.~Veselov, \emph{Phys.
Rev. D} \textbf{79}, 034003 (2009)

%

\end{thebibliography}
\end{document}